# High-resolution imaging and spectroscopy in the visible from large ground-based telescopes with natural guide stars.

Craig Mackay[a, *], Tim D. Staley[a], David King[a], Frank Suess[a] and Keith Weller[a]
[a]Institute of Astronomy, University of Cambridge, Madingley Road, Cambridge, CB3 0HA, UK;

**ABSTRACT**

Near-diffraction limited imaging and spectroscopy in the visible on large (8-10 meter) class telescopes has proved to be beyond the capabilities of current adaptive optics technologies, even when using laser guide stars. The need for high resolution visible imaging in any part of the sky suggests that a rather different approach is needed. This paper describes the results of simulations, experiments and astronomical observations that show that a combination of low order adaptive optic correction using a 4-field curvature sensor and fast Lucky Imaging strategies with a photon counting CCD camera systems should deliver 20-25 milliarcsecond resolution in the visible with reference stars as faint as 18.5 magnitude in I band on large telescopes. Such an instrument may be used to feed an integral field spectrograph efficiently using configurations that will also be described.

## 1. INTRODUCTION

It is nearly 20 years since the launch of the Hubble Space Telescope (HST), an instrument which transformed our view of the universe by providing a tenfold step change in the angular resolution of visible and near infrared astronomical images. Although the many larger ground-based telescopes have a diffraction limit much smaller than the ~0.12 arc sec resolution images routinely delivered by HST, progress in achieving better resolution has been slow, expensive and of rather mixed success. The degradation in image quality caused by the effects of atmospheric turbulence has been more difficult to overcome than anticipated. Many major observatories have invested heavily in adaptive optic systems and good high-resolution images have been obtained particularly in the near infrared where the effects of turbulence are much less severe. In the visible, however, it is still the case that no adaptive optic system has achieved HST resolution on an HST size (2.5 m diameter) telescope. This lack of success inevitably calls into question the ambitious plans to correct the turbulent wavefronts entering the next generation of very large telescopes. There is, however, a rather different approach to ground-based visible imaging which is already delivering HST resolution on HST size telescopes. Originally studied by Fried[2] following earlier work by Hufnagel[3] it relies on taking a large number of images with exposure time short enough to freeze image motion due to atmospheric turbulence. A reference star in the field allows the relative sharpness of each image to be determined. The sharpest images are selected then shifted and added to provide a high resolution composite image. On a good site under typical seeing conditions it is possible to produce HST resolution routinely on 2.5 m class telescopes. An example is shown in Figure 1. The statistics of turbulence makes the technique difficult to apply directly to use on larger telescopes. The technique depends on the use of high-speed photon counting imaging cameras. We have had considerable success with using the electron multiplying CCD detectors manufactured by E2V Technologies Ltd (Chelmsford, UK). With careful design they can be made very close to perfect, noiseless, photon counting imaging systems that are critical to the success of this work.

*cdm <at> ast.cam.ac.uk

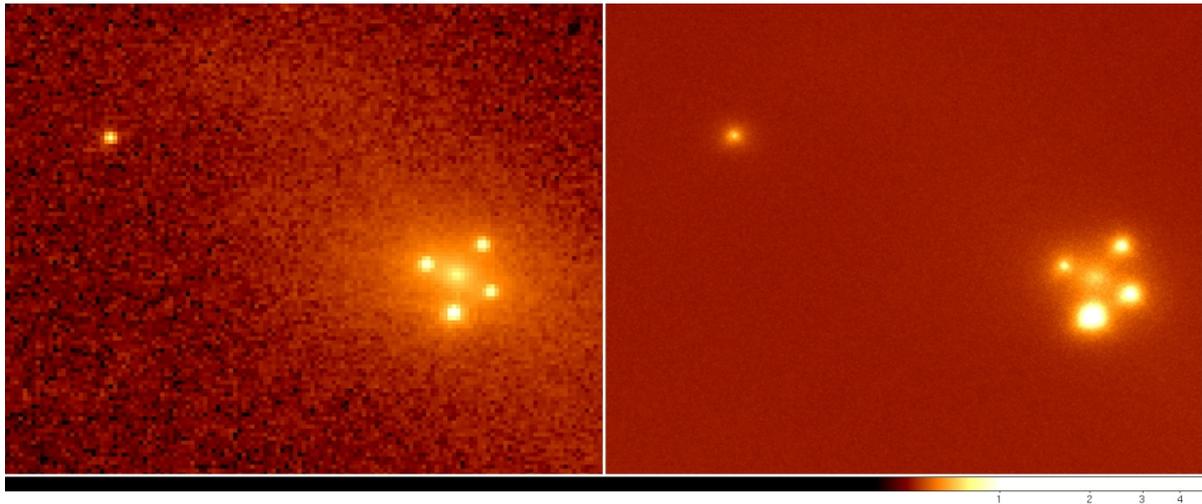

Figure 1: Images taken of the Einstein cross, which shows four gravitationally lensed images of a distant quasar magnified by the core of a relatively nearby bright galaxy. The left-hand image was taken with the Hubble Space Telescope Advanced Camera and the right hand image taken with the Cambridge LuckyCam attached to the NOT 2.5 m telescope on La Palma.

The majority of instruments in use and being designed to give a high order of wavefront correction on large telescopes use Shack-Hartmann wavefront sensors. These work by forming an array of separated images of a reference star from well-defined areas of the telescope pupil. By tracking the movement of each of these images as they are deflected by corrugations in the wavefront entering the telescope it is possible to compute the compensating deflections that will be needed to correct the wavefront. Those corrections are used to drive a deformable mirror very rapidly so that the wavefront corrugations are substantially eliminated. The number of sub images used (typically several per square metre telescope aperture) and the readout rate to be used (typically several hundred Hz frame rate) determine the brightness of the reference star. This is typically 12-13 magnitude, a level that ensures there is only a of very small chance of finding such a reference star close to a random science object. More recently, laser guide stars are being used to improve sky coverage. In practice they create their own problems. Even neglecting cone effect and other aberrations, laser guide images are seldom smaller than about one arc sec in diameter. They cannot generally be centred to better than one tenth of that size and so are unlikely to deliver angular resolution significantly better than that of HST.

It is of the greatest importance to astronomers to be able to work in the visible since that is where most of our knowledge of the physical universe has been gained. It is in the visible that we are best able to measure stellar types and metallicities, and where we know most about the astrophysics of hot gases in nebulae. Wavefront correction will always be much harder in the visible in the near infrared since it scales as ~1/wavelength[1,2]. However the rewards in terms of enhanced resolution of working in the visible rather than the infrared very great indeed. An ideal system would work in the visible on current 8-10 m class telescopes to deliver 20-25 milliarcsecond angular resolution images that could feed efficiently a high-resolution spectrograph system. Such a system should be able to work with faint natural guide targets so that sky coverage approaches 100%, and it must deliver an isoplanatic patch size that allows reliable measurements of the characteristics and shapes of both guide and science targets. The very small isoplanatic patch size delivered by conventional Shack-Hartmann based AO systems is a major problem. The point spread function (PSF) varies rapidly across the field making photometry and astrometry extremely difficult. The I-band isoplanatic patch size with conventional AO is only a few arcsec in diameter.

This paper will discuss some of these problems and show how they might be overcome in a practical system, a prototype of which that has already delivered the highest resolution images ever taken in the visible on ground-based telescope with an angular resolution of about 35 milliarcseconds.

## 2. THE CHALLENGE OF THE SKY

Bright reference objects are very scarce, particularly at high Galactic latitudes. Counts of stars have been presented by Simons[1] in R-band. Using a mean colour index of R-I of 1.5 mags, stars with I=20.5 arc found at around one per square arc minute at high Galactic latitudes, I=19.3 at 60°, and I=18 at 40°. When considering reference objects at these magnitudes, it is important to remember that many faint galaxies have compact cores which, if small enough, could also be used as reference objects. At I=19, galaxies are more common than stars at the Galactic poles[5]. Although I=19 might seem rather faint, on an 8m diameter telescope such as star can give a detected photon rate of over 4000 photons per second in I band and in excess of 10,000 photons per second using a standard thinned (back illuminated) CCD with a filter passing everything longward of 600 nm.

It is clear that these very faint reference stars could not be used with a Shack-Hartmann sensor with many hundreds of elements and running at several hundred frames per second. If we are to reduce the effects of turbulence on our images then we need to take a very different approach.

Lucky imaging techniques have been described in detail by Baldwin et al [4] and references therein, as they apply to HST size telescopes in I band. The probability of recording a high resolution image under these circumstances is typically 10-30%. With larger telescopes the probability diminishes rapidly. On a 5 m telescope, for example the same seeing conditions would give a probability of $10^{-4}$ or 0.8 %, and on 10 m telescope the probability becomes quite negligible for two reasons. Firstly, the number of turbulent cells increases with the area of the telescope mirror dramatically reducing the chance of a lucky exposure. Secondly, there is now turbulent power on large scales. The power spectrum of atmospheric turbulence approximately follows the predictions of Kolmogorov [9] and is dominated by turbulent power on the largest scales. Lucky imaging works when the average number of turbulent cells across the diameter of the telescope is in the range of 6-18 [4,10]. The only way the lucky imaging might work on a large diameter telescope is if a significant part of the turbulent power can be removed so that the turbulent cell sizes are effectively increased in proportion to the increased mirror diameters.

## 3. ATMOSPHERIC TURBULENCE CHARACTERISTICS

Atmospheric turbulence distorts the wavefront entering the telescope so introducing significant phase errors that smear out the final image. Although the turbulence that affects astronomical imaging is generated by interactions between different layers of airflow at altitudes of typically 10 km, the stirring motion than many imagine continues in fact stops within a few turning times. We can see that by looking at high altitude contrails left by highflying aircraft. At their altitude, also typically ~10 km, complex patterns are generated rapidly that can persist for many minutes. These represent regions of slightly different temperature that become frozen into the airflow. The same is true for astronomical turbulence: the turbulent pattering is largely frozen into the airflow and the timescales over which the turbulence effects on the image change is comparable to the wind crossing time of the telescope. This means that, although the level of turbulence is higher across a larger diameter telescope, it effectively changes proportionately more slowly.

It is important to understand one essential difference between the Shack-Hartmann approach to wavefront correction and the lucky imaging approach. With typical sub image scales corresponding to ~50cm in the telescope pupil plane it is essential that the shack-Hartmann sensor readout repeats on a timescale significantly smaller than the wind crossing time for a 50 cm aperture. With lucky imaging, the timescales can be very much longer. Imagine the (very artificial) situation with a completely flat wavefront apart from one small area. Once that distorted area is within the pupil, the image in the focal plane will not change while that area is blown across the pupil, and until it has moved outside the pupil. Therefore a single lucky exposure equal to the wind crossing time of the telescope would give a high resolution image. In reality, the wavefront is much more complex, but the principal point here is that the timescale on which lucky images evolve are much longer than the timescales over which a Shack-Hartmann sensor image changes. On the Paranal site in Chile of the VLT, median wind speeds are 6-7 km/s so the wind crossing time for an 8.2m telescope is typically > 1 sec. From this we see that the largest scales of turbulence cross the telescope relatively slowly. This suggests the possibility that the largest scales of turbulence could be determined and corrected for on much longer timescales and so an appropriate wavefront sensor could be operated with much longer exposure times, longer by a factor ~ (telescope diameter/0.5m), a factor of of 16/20 for 8/10m telescope diameters. This in turn would allow much fainter reference stars to be used.

It is interesting to consider the relative turbulent power on different scales predicted by standard Kolmogorov turbulence models. Noll[11] has examined the contribution of atmospheric turbulence when broken down by Zernike coefficient. He shows that eliminating tip-tilt errors reduces the power to 13.3% of its original levels, that removing the additional three Zernike modes after tip tilt correction (Z4-Z6) removes 93.1% and removing the next four modes (up to Z10) additionally leaves only 1.6% of the original power. It is clear from this that the extraordinary effort needed to remove higher powers will only give a relatively modest improvement in image sharpness.

A further point that is important to consider is exactly what our goals should be in terms of improving the sharpness of astronomical images from ground-based telescopes. It is traditional for people working in adaptive optics to seek after the highest possible Strehl ratio and there are indeed a small number of programs were this is likely to be critical. A good example here is the search for extrasolar planets which is often principally limited by scattered light from the star around which planets are being sought. In this case the highest Strehl ratio possible is critical to the success of the work. However for many other studies the ultimate Strehl ratio is much less important. It is the ability to resolve objects with confidence and to observe them with a well-defined and quantifiable point spread function even if that point spread function does not correspond to a particularly high Strehl ratio. The consistency of point spread function is much more important since that may be used with deconvolution methods, for example, to establish meaningfully whether objects are resolved and what is their proper structure. In practice, images with a Strehl ratio as low as 0.1 are perceived visually as being sharp. Much higher Strehl ratios produce images that are perceived visually as being only marginally sharper. Yet it is the drive for these higher Strehl ratios that substantially drives the choice of technologies that are so difficult to implement.

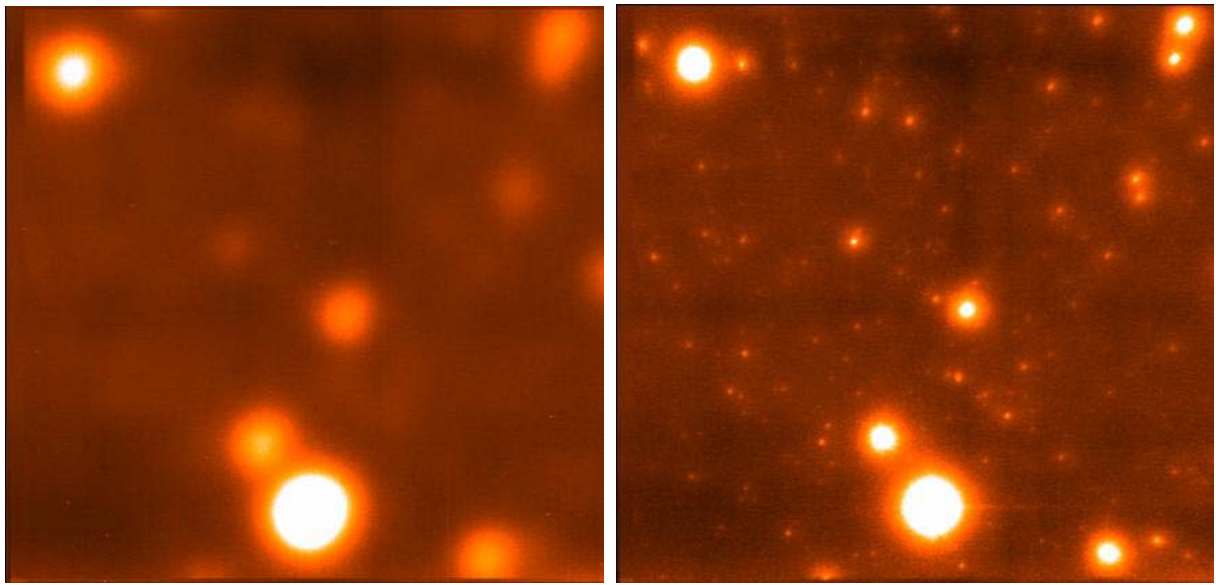

Figure 2: The Globular cluster M13 as imaged conventionally by the Palomar 200 inch telescope (left), followed by M13 as imaged with the Lucky Camera behind an adaptive optics system on the Palomar 200 inch telescope (right). The natural seeing was about 0.65 arcsec, and the Lucky/AO image has a resolution of about 35 milliarcsecs or about three times that of the Hubble Space Telescope. This image is the highest resolution image ever taken with any optical or infrared telescope on the ground or in space..

In order to test whether the idea of removing the largest scales of turbulence to allow lucky imaging to succeed, we attached one of the high-speed electron multiplying, photon counting CCD detector systems from PixCellent Imaging Ltd. onto the Palomar 5 m telescope, behind the Palomar adaptive optics system PALMAO [12]. Conventional lucky imaging would be predicted simply not to work on a telescope of this size, giving a negligible fraction of sharp images even under the best seeing conditions. The PALMAO instrument is a relatively low order adaptive optics system with 241 active actuators and 16 x 16 sub aperture Shack-Hartmann wavefront sensor. The results obtained were extremely encouraging. Images with a limiting resolution of about 35 milliarcseconds were obtained in I band and are the highest resolution images ever taken in the visible or near infrared anywhere. Even with 50% image selection the full width at half maximum of star images was approximately35 milliarcseconds (Figure 2). It was clear that are similar instrument on a larger telescope would undoubtedly have produced images with yet high angular resolution in the visible. The principal disadvantage of the PALMAO instrument was that it needs the usual relatively bright reference object that is needed by Shack-Hartmann wavefront sensors.

## 4. CURVATURE WAVEFRONT SENSORS

Shack-Hartmann sensors are the most common type used for adaptive optics wavefront sensing, but Racine [6] has looked at the limiting sensitivities of wavefront sensors currently deployed on telescopes and found that the limiting sensitivity old curvature sensors[7] are typically 2.5 magnitudes fainter than Shack-Hartmann sensors although this difference is progressively reduced with higher order correction requirements. One of the best-known and most successful curvature sensors is Hokupa'a, a natural guide star, curvature-sensing adaptive optics system built by the University of Hawaii [7] with 36 elements. This used a flexible bimorph mirror in order to image on either side of the pupil plane driven with a loudspeaker voice coil between the two positions. The implementation of the curvature sensor concept was very demanding when this work was done. Although by modern standards this is a rather low order adaptive optic sensor it was remarkably successful in delivering significant image resolution enhancement. Recent developments, particularly in detector technology, allow a much simpler and much more efficient approach.

Guyon et al [8,15] looked in detail at the design of curvature wavefront sensors. Curvature sensors have advantages in that they need many fewer detector elements in the wavefront sensor and can achieve higher Strehl ratios with the same number of actuators. The sensitivity advantage of curvature sensors which diminishes for higher order corrections may be avoided by imaging four planes, two on either side of the pupil plane to accommodate the intrinsic nonlinearity of the system. They suggest that a more complicated bimorph scanning pattering can provide these four image planes but it is inevitable that a significant fraction of the total time used to sense the wavefront is occupied by scanning from one plane to the next. We propose a rather different optical arrangement where the four planes are imaged simultaneously onto a single photon counting electron multiplying CCD. The light from the reference star is split four ways and then reimaged on to a single CCD detector as 4 separate images. With an efficient photon counting detector there is no disadvantage in reading out at relatively high speeds since this minimises the probability of having more than one photon per pixel per frame and thereby risking saturating the detector. Consecutive frames may be combined to improve the signal-to-noise of the wavefront determination. The numbers of frames combined may be adjusted in response to the atmospheric conditions and the brightness of the reference object. Another major advantages of curvature sensors over Shack-Hartmann sensors is that the former are able to work at much lower light levels. The minimum signal that a Shack-Hartmann wavefront sensor can tolerate corresponds to approximately 100 photons detected per lenslet per frame time. Below this level, all the sensor elements cease to function simultaneously, and no correction information is available. However, with curvature sensors even when higher orders of correction are no longer possible because of lack of light from the reference star it is still possible to derive low order correction information. A further advantage of the designs developed by Guyon et al [8,15] is that the curvature sensors are relatively achromatic and so allow a very wide bandpass to be used, thereby maintaining the sensitivity advantage of this technology.

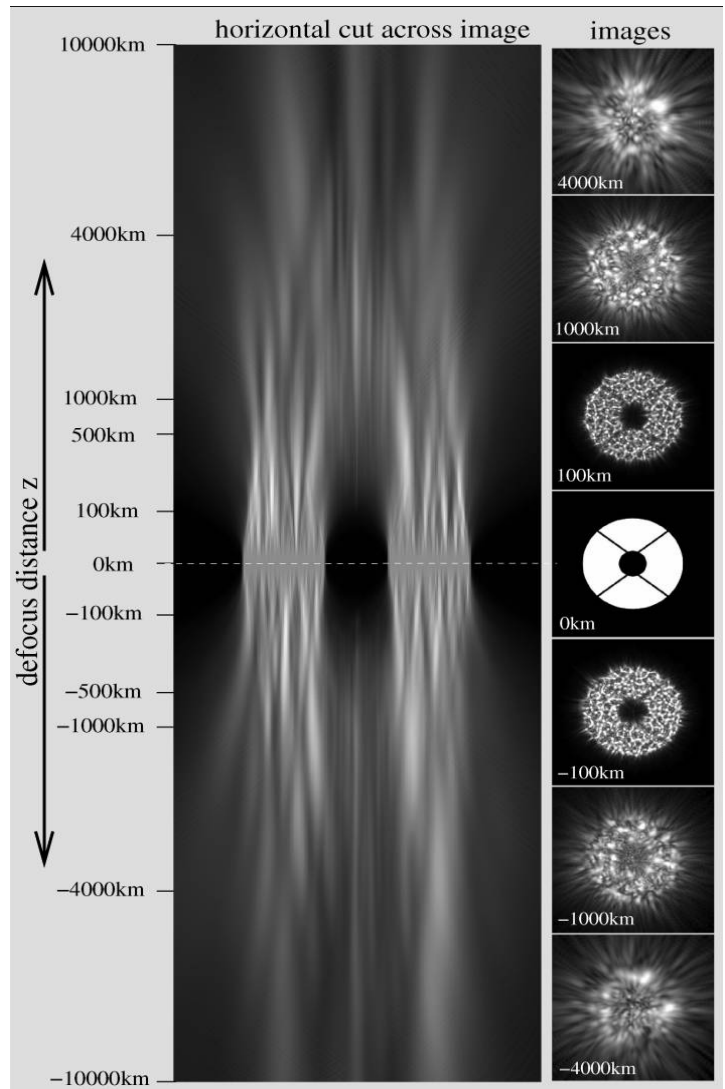

Figure 3: Simulation of the propagation of light through the pupil of starlight affected by atmospheric turbulence (figure 1 from Guyon et al [15])

Guyon et al [8] have simulated the propagation of a turbulent wavefront through the pupil of the telescope. Figure 3 (figure 1 from Guyon et al [15]) shows how the detected wavefronts change in characteristic on either side of the pupil plane. At small displacements the otherwise uniform illumination of the pupil starts to break up into speckles corresponding to the smallest turbulent scales present. At the larger displacements the structure of the pupil images is dominated by the larger scales of turbulence. One weakness of many existing curvature sensor systems is that they are relatively inefficient at detecting tip-tilt errors. This is because these produce lateral shifts in the images taken on either side of the pupil. By differencing them only a small edge effectively contribute to the area signal. By detecting the near-pupil images with a relatively high resolution detector it is possible to use the speckle structure in these images to give a much more accurate measurement of tip tilt even at relatively low photon rates. Guyon et al [15] have also shown that the accuracy with which the wavefront errors may be determined is greatly improved by using four image planes rather than the more conventional two since the propagation through the pupil is non-linear. They further demonstrate that this approach greatly improves the achromaticity of the technique. The data shown in figure 3 correspond to simulations using a 0.4-0.8 μ bandpass. Again, the ability to use such wide band passes is extremely attractive as it further improves the sensitivity of the curvature sensor.

# 5. CURVATURE SENSOR SIMULATIONS

The work of Guyon et al [8] was particularly targeted at the development of relatively high order curvature sensors in order to achieve the highest Strehl ratios. Our studies have emphasised the importance of being able to work with the faintest possible reference stars in order to give the widest coverage. In order to establish just how faint a reference star it is possible to work with in this configuration we have simulated the performance of the instrument described above. After the light passes through the telescope focal plane it is collimated onto a deformable mirror at a pupil plane and then reimaged. A small pickoff mirror passes light from the reference object to a collimating lens, and then the light is split into four separate beams. Reimaging optics form images on either side of the pupil plane which are then placed side-by-side on the photon counting electron multiplying CCD detector. In our simulations the location of the four near-pupil plane images relative to the true pupil plane is adjustable.

Our instrument will use electron multiplying CCDs for both the science beam and curvature sensor beam. They will be run in synchronism except that the curvature sensor will be run at four times the frame rate, using only a 200 x 1024 pixel patch of a 1024 x 1024 EMCCD [13]. The curvature sensor data may then be averaged both spatially and temporally depending on conditions such as seeing and windspeed. The four out-of-pupil images are then fitted to a simple wavefront propagation model using phase retrieval methods. The derived wavefront phase error referenced to the pupil plane (in which a deformable MEMS mirror[14] is placed) is used directly to drive the MEMS device.

Our simulations predict that we should be able to increase the effective $r_0$ by removing the largest scales of turbulent power so that the number of turbulent cells across the 8.2m VLT mirror is the same as we typically see across a 2.5m mirror without any low-order correction. This is predicted to work with reference stars as faint as I = 19. The wavefront measured also allows the Lucky Image selection to be made from the curvature data. We will be able to confirm this by testing the system with brighter reference stars but using a 50% reflecting pick-off spot so that we still see the reference star in the science beam and can check that the wavefront selection criteria are being properly established.

# 6. HIGH RESOLUTION INTEGRAL FIELD SPECTROSCOPY

The techniques described are intended for direct imaging applications. It is also possible to use the low-order AO correction subsystem to feed a high resolution spectrograph using an integral field unit. These IFUs consiste of a close-packed array of tiny lenses each of which concentrates the light that falls on it into a single multi-mode fibre. The output ends of the fibres are placed into a line to mimic the slit of a conventional slit spectrograph, and dispersed. In parallel a second IFU is placed at the position of a reference star image, and again the fibre outputs are lined up but here not dispersed. The light from the reference IFU is checked. A sharp frame occurs when the reference star illuminates only one fibre principally, and the specific fibre shows how this sharp frame is aligned relative to other frames. The spectral image is then selected, shifted and added as before. The lenslets may be made very small with an appropriate optical feed so that high spatial resolution may be achieved.

# 7. CONCLUSIONS

We have described a design of a novel strategy of achieving near-diffraction limited imaging and spectroscopy on 8-10m class telescopes in the visible for the first time. Key technologies are the electron multiplying CCDs that allow photon-counting performance at high frame rates, and a new approach to wavefront sensor design employing a four extra-pupil curvature sensor structure. It has the potential to revolutionise ground-based imaging in the visible and give astronomers the capacity to study objects with unrivalled resolution.